\documentclass{iaus}
\usepackage{graphicx,natbib}
\def\aj{AJ}%
\def\apj{ApJ}%
\def\apjl{ApJ}%
\def\aap{A\&A}%
\def\mnras{MNRAS}%
\def\pasp{PASP}%
\title[ICL from SDSS stacking] %% give here short title %%
{Statistical Properties of the IntraCluster Light from SDSS Image Stacking}

\author[S. Zibetti] %% give here short author list %%
{Stefano Zibetti$^1$%
  \thanks{Current affiliation: Max-Planck-Institut f\"ur Astronomie -
    K\"onigstuhl 17, 69117 Heidelberg - Germany} }

\affiliation{$^1$Max-Planck-Institut f\"ur extraterrestrische Physik -
Postfach 1312 - D-85741 Germany}

\pubyear{2007}
\volume{244}  %% insert here IAU Symposium No.
\pagerange{209--218}
\date{July 26th, 2007 and in revised form August 24th, 2007}
\jname{Dark Galaxies \& Lost Baryons}
\editors{J. I. Davies \& M. J. Disney, eds.}
\begin{document}

\maketitle

\begin{abstract}
The presence of a diffuse stellar component in galaxy clusters has
been established by a number of observational works in recent
years. In this contribution I summarize our results \citep{SZ+05}
obtained by stacking SDSS images of 683 clusters, selected with the
maxBCG algorithm at $0.2<z<0.3$. Thanks to our large sample
($\gtrsim30$ times larger than any other sample of individually
observed clusters so far) and the advantages of image stacking applied
to SDSS images, we are able to measure the {\em systematic} properties
of the intracluster light (ICL) with very high accuracy.

We find that the average surface brightness of the ICL ranges between
26 and 32 mag arcsec$^{-2}$, and constantly declines from 70 kpc
cluster-centric distance (i.e. distance from the BCG) to 700 kpc.
Interestingly, the {\em fraction} of diffuse light over the total
light (including galaxies), monotonically declines from $\sim50$ to
$\lesssim 5$\% over the same range of distances, thus showing that the
ICL is more easily produced close to the bottom of a cluster's
potential well. On the other hand, clusters lacking a bright BCG
hardly build up a large amount of intracluster stellar component.  The
link between the growth of the BCG and the ICL is also suggested by
the strong degree of alignment between these two components which is
observed in clusters where the BCG displays a significant elongation.
With the additional fact that the colors of the ICL are consistent
with those of galaxies, all this appears to be evidence for
intracluster stars being stripped from galaxies that suffer very
strong tidal interactions in the center of clusters and eventually
merge into the BCG.

Our measurements also show that intracluster stars are a minor
component of a cluster's baryonic budget, representing only $\sim10\%$
of the total optical emission within 500 kpc.

Finally, we discuss some open issues that emerge from a comparison of
the present results with other observations and recent theoretical
modeling.


%% add here a maximum of 10 keywords, to be taken form the file <Keywords.txt>
\end{abstract}

\firstsection % if your document starts with a section,
              % remove some space above using this command.
\section{Introduction}

The existence of a diffuse stellar component in galaxy clusters is now
a well established observational fact. After the early claim by
\cite{1951PASP...63...61Z} and pioneering work in the 1970's
\citep{welch_sastry_71,melnick_white_hoessel77}, the advent of
high-sensitivity panoramic CCD detectors allowed the unambiguous
detection of intracluster light (ICL) in nearby clusters at very faint
surface brightness levels, from $\sim26$ down to $\sim30$mag$_R$
arcsec$^{-2}$
\citep{bernstein+95,gonzalez+00,2005ApJ...618..195G,feldmeier+02,feldmeier+04}.
More recently, \cite{2007AJ....134..466K} have reported observations
of the ICL in clusters up to $z\sim0.3$.

Intracluster stellar populations have gained relevance in recent years
as they can provide extremely useful constraints in a number of
cosmological issues, among those: the history of galaxy interactions
and mergers in dense environments, the baryonic budget of close systems
\citep[][this conference]{2007arXiv0705.1726G}, the evolution of the
stellar mass function of luminous red galaxies
\citep{2006ApJ...652L..89M}.

Despite the growing number of observations fundamental questions about
the nature and the origin of the ICL still remain open. The first
issue concerns the very definition of the ICL. From a dynamical point
of view one should consider as ICL only the light radiated from stars
which are {\it not} bound to any individual galaxy, rather free
floating in the cluster potential. Such a strict definition is not
applicable to photometric measurements like those mentioned above,
unless one believes parametric surface brightness decomposition based
on analytical functions. To access the required dynamical information
the observation of intracluster planetary nebulae (ICPNe) has been
proposed and applied to nearby clusters
\citep[e.g.][]{arnaboldi+96,feldmeier_ciardullo+04}. These kinds of
studies have demonstrated that the genuine intracluster stellar
component and the extended halos of bright (elliptical) galaxies are
often overlapping and can be hardly distinguished based on broad band
photometry alone. \cite{2004ApJ...614L..33A} convincingly show that
the ``binding'' state of ICPNe that trace the diffuse stellar
population critically depends on local density (cluster core or
periphery), on the mass of neighboring galaxies and on the dynamical
state of the cluster and its sub-components. All photometric studies
of the ICL have to {\it define} criteria, like surface brightness
thresholds and/or profile decompositions, to isolate the ICL from the
rest of the optical cluster light, in a way that does not need to be
consistent with dynamical definitions. Therefore attention should be
paid at possible differences between ``photometric'' and ``dynamical''
ICL, for instance when comparing photometric studies with simulations.
Also, different definitions in a number of recent studies can give
rise to confusion in the quantitative results. The apparent general
agreement between the results published in literature should be
regarded with some care before deriving firm conclusions, as I will
discuss later.\\
The way the ICL is defined and measured is also very relevant for
studies that attempt at computing the total stellar budget in clusters
\citep[e.g.][this conference]{2007arXiv0705.1726G}, as the light in
galaxies and in the IC component must be measured consistently, for
instance by applying the same SB thresholds to both components.

Although there is general consensus on the IC stars being formed in
galaxies and subsequently being scattered into the intergalactic
space, it is still not completely clear which are the main mechanisms
of production. A number of theoretical studies, ranging from
N-body+SPH cosmological simulations
\citep[e.g.][]{2005MNRAS.357..478S,2007MNRAS.377....2M}, to
dissipationless N-body simulations
\citep{2006ApJ...648..936R,2007astro.ph..3374C}, to pure analytic
models \citep{2007astro.ph..3004P}, give somehow contrasting
explanations for the origin of the ICL, from the scatter of stars
during the mergers of galaxies into the central cluster galaxy to
tidal stripping and/or disruption of dwarf satellites. Observations
show all these mechanisms to be at work to some extent.
\cite{2007A&A...468..815G} clearly show that a large number of
intracluster stars are being ``created'' in the ongoing merger of the
two brightest ellipticals in the core of the Coma cluster. On the
other hand, the wealth of tidal structures ubiquitously observed in
nearby clusters \citep[as seen for instance in the spectacular image
of Virgo published by][]{2005ApJ...631L..41M} testifies that IC stars
can be ``created'' all over the cluster environment. Given this large
phenomenological variance, a systematic study of the properties of the
ICL in a broad variety of clusters drawn from a large statistical
sample is required in order to derive cosmologically meaningful
constraints to the physics of cluster and galaxy formation. This was
exactly the main goal of the study that we published in 2005
\citep{SZ+05} and whose results are the focus of this contribution.

In the following I will first (Section \ref{sec:sample_photo})
introduce the sample and the image stacking technique we have adopted
in our analysis. In Sec. \ref{sec:results} I will summarize the main
results.  Section \ref{sec:interpretation} is devoted to a short
discussion about the physical mechanisms of ICL production that are
suggested by the present results. I will then discuss some critical
issues that need to be clarified and addressed by future observations
and modeling (Sec. \ref{sec:open}) and finally give some conclusions
(Sec. \ref{sec:conclusions}). A fully detailed description of data
processing and analysis is beyond the scope of this contribution and
can be found in \cite{SZ+05}.

Throughout we adopt the standard cosmological parameters
$\Omega_{tot}=1.0$, $\Omega_{\Lambda}=0.7$, $h=0.7$.

\section{Sample and photometric analysis}\label{sec:sample_photo}
As already mentioned above, the diffuse ICL has very low typical
surface brightness $\mu_R>26$ up to
30--31$~{\mathrm{mag~arcsec}^{-2}}$, which is of the order of
$10^{-3}$ or less times the surface brightness of a dark night sky. At
these extremely faint levels not only sensitivity is an issue, but
also and more crucial are systematic effects due to residual from flat
field corrections, scattered light, internal reflections in the camera
and fringing (in the red bands). These problems hamper the
observations of ICL in individual clusters and make such observations
very expensive in terms of telescope time and strategy \citep[for
adopted solutions see
e.g.][]{2005ApJ...618..195G,2007AJ....134..466K}.  This has prevented
to collect observations of the ICL in statistically large samples
($N\gtrsim30$) of clusters so far.

The SDSS, with its large homogeneous drift-scan imaging coverage in 5
bands, offers the opportunity to tackle this observational problem
with a completely alternative statistical approach. Instead of
analyzing each cluster individually, it is possible to stack several
hundreds images of clusters after masking out the light of all
galaxies. In this way we can not only enhance the sensitivity to the
levels required to study the ICL, but also the systematic ``defects''
mentioned above average out in the composite image.

For this experiment we select cluster candidates from
$\sim1500$~deg$^2$ of SDSS-DR1 as given in an unpublished preliminary
version of the maxBCG cluster catalog
\citep{2007ApJ...660..239K,2007ApJ...660..221K} kindly provided by Jim
Annis. We selected clusters in the redshift range between 0.2 and 0.3
as a compromise between a sufficiently large image coverage of the
cluster region over a single SDSS frame and minimal cosmological SB
dimming. Before stacking, each image was centered on the brightest
cluster galaxy (BCG), rescaled on a common physical size and
photometric calibration, and checked against photometric defects
(bright stars, evident blooming or scattered light). The final sample
is composed of 683 clusters. Using the number of red sequence galaxies
provided by the maxBCG catalog and the empirical relation between this
number and $R_{200}$ \citep{2005ApJ...633..122H}, we can derive a
rough estimate of the masses $M_{200}$ using the formulae of spherical
collapse theory. Our sample includes objects ranging from a few
$10^{13}$ up to $5~10^{14}~h_{70}^{-1}\mathrm{M_\odot}$, that is from
(rich) groups to quite massive clusters\footnote{These estimates were
  not given in \cite{SZ+05}.}. Also interesting for following analysis
is the fact that we span a large range in BCG luminosity, namely from
$\sim -22$ to $\sim -24$ mag$_r$ (corrected to rest-frame $z=0$).

We chose to define as ICL all the light emitted from regions having a
surface brightness fainter than 25 mag arcsec$^{-2}$ ($r$-band in the
$z=0.25$ observed frame). All galaxies, except the BCG, are masked to
this extent \citep[but see][for details about masks and relative
corrections]{SZ+05}. The BCG is left unmasked in the stacking process
and is treated separately as its halo contribution to the ICL is
debated, although in many respects this is just matter of semantic.

\section{Results}\label{sec:results}
\begin{figure}
\includegraphics[width=0.5\textwidth]{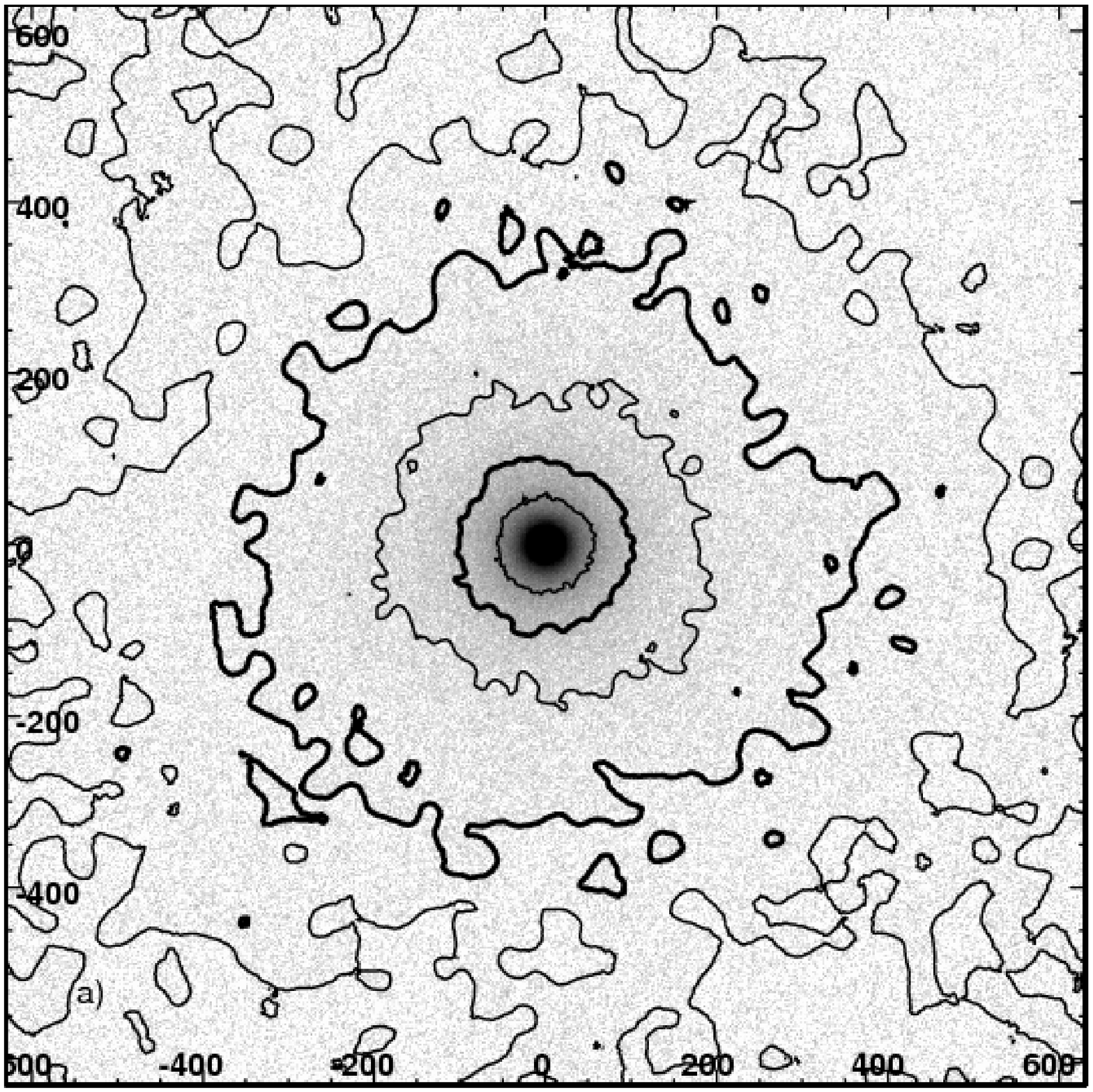} 
\includegraphics[width=0.5\textwidth]{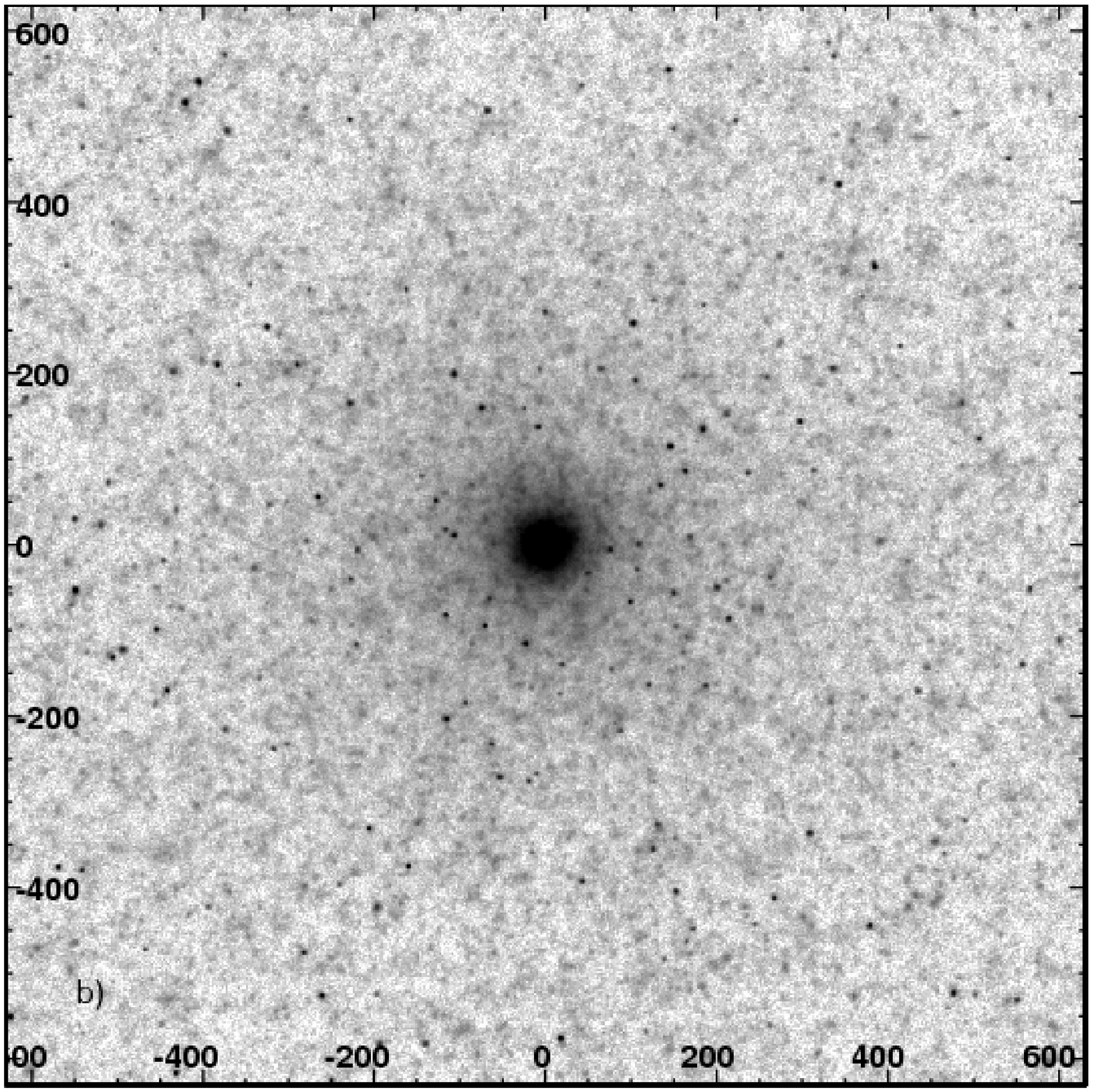}
\caption{The $r+i$ composite images resulting from the stacking of the
  main sample: the diffuse component plus BCG is in panel a), the
  total light in panel b). The same logarithmic grey scale is adopted
  in both images. Side-scale tickmarks display the distance in kpc
  from the center.  Isophotal contours corresponding to
  $\mu_{(r+i),0.25}$ of 26, 27, 28, 29 and 30 mag~arcsec$^{-2}$ for
  the diffuse component are overplotted on panel a). Smoothing kernels
  of 3, 7, 11, 17, and 21 pixels respectively are used. Corresponding
  SB values in $r$-band are $\sim 0.3$ mag
  brighter.}\label{fig:stackimages}
\end{figure}

The pictorial result of the stacking for the whole sample of 683
clusters is presented in Fig. \ref{fig:stackimages}. Panel (a) shows
the stack image where all galaxies except the BCG are masked out and
represent the \emph{average} distribution of light in the BCG and
diffuse component. The faintest isophotes represent $\sim 30.3$~mag
arcsec$^{-2}$ ($r$-band observed frame) and extend to distances of
$\gtrsim 600$~kpc. For comparison, panel (b) shows the stack image
where only bright foreground sources have been masked, such that the
total cluster light is represented.
\begin{figure}
\centerline{\includegraphics[width=\textwidth]{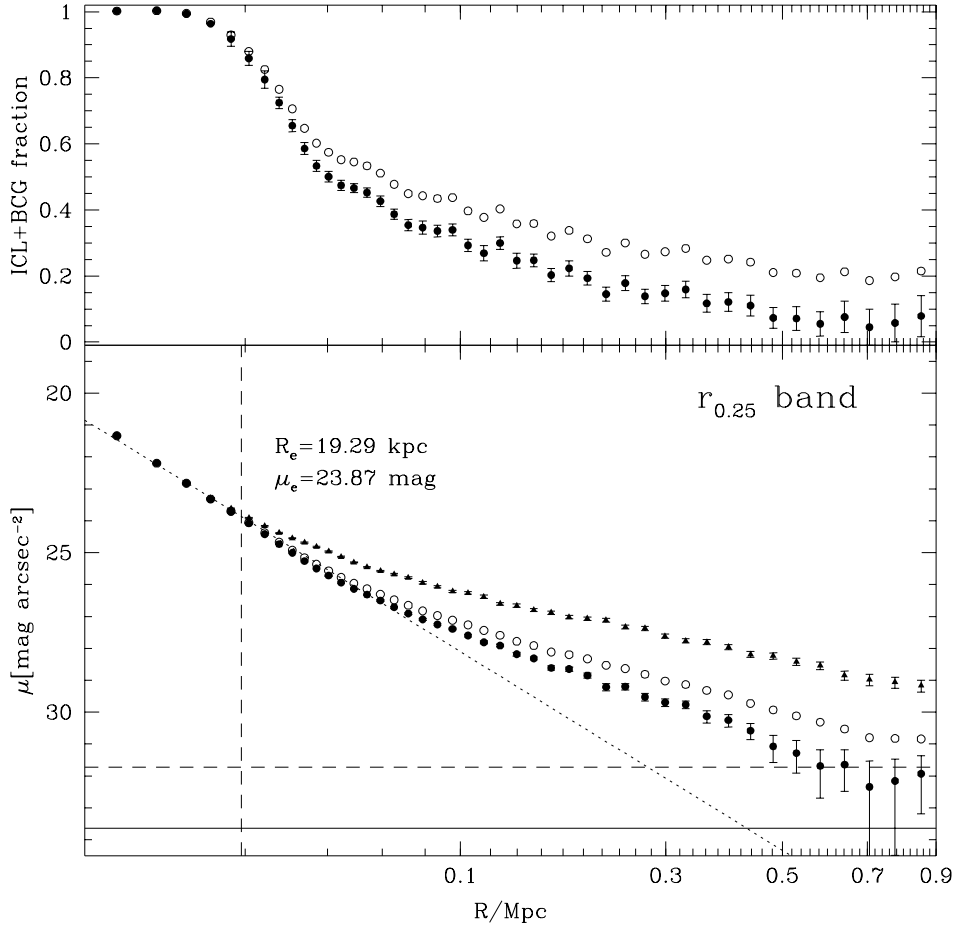}}
\caption{The surface brightness profiles and the local ratio of
  \textit{ICL+BCG} and uncorrected \textit{diffuse} light to the {\it
    total} cluster light in the $r$ band for the complete sample of
  683 clusters.The $R$ axis is linear in $R^{1/4}$. Bottom panel: the
  SB is expressed in mag arcsec$^{-2}$ in the $z=0.25$ observer frame.
  Triangles with error bars represent the total cluster light, open
  circles the diffuse light (including the BCG) as directly measured
  from the stacked images. Filled circles with error-bars display the
  SB of the {\it ICL+BCG}, corrected for masking incompleteness.
  Horizontal dashed and solid lines display the SB corresponding to
  the 1-$\sigma$ uncertainties on the background determination for the
  total light and for the {\it ICL+BCG} respectively. The dotted lines
  represent the best de Vaucouleurs fits to the inner regions: the
  effective radius of the best fitting model is indicated by the
  vertical dashed line and the corresponding parameters are reported
  nearby. Top panel: the local ratio of {\it ICL+BCG} (filled dots
  with error-bars) and uncorrected \textit{diffuse} light (open
  circles) to {\it total} cluster light.}\label{fig:r_profile}
\end{figure}
From stacked images like these we extracted azimuthally averaged
surface brightness (SB) profile in each band and for different
subsamples of clusters. An example of such SB profiles is reported in
Fig. \ref{fig:r_profile}.  Three main results can be extracted from
these profiles.
\begin{itemize}
\item{There is a diffuse luminous component (ICL+BCG) that extends to
    cluster-centric distance of $\sim700$~kpc at the limiting SB of
    $\sim32$~mag~arcsec$^{-2}$}.
\item{The SB of the diffuse component declines with distance more
    steeply than the total light of the cluster. Its contribution
    declines from roughly 30\% of the total at 100 kpc to almost 0 at
    700 kpc.}
\item{The diffuse light is dominated by the DeVaucouleurs component of
    the BCG in the inner 70--80 kpc. Outside this radius the diffuse
    light profile significantly flattens and the light is dominated by
    galaxies other than the BCG (the fraction of ICL+BCG over total
    drops below 50\%). If we assume that light traces mass, this
    implies that the flatter outer diffuse component lives in regions
    where the global cluster potential dominates over the potential of
    the self-bound central galaxy, and therefore can be dubbed as
    genuine ICL.}
\end{itemize}
It is worth noting at this point that, with respect to other studies
of the ICL in individual clusters, the stacking analysis is extremely
powerful in terms of depth and spatial extensions. In comparison to
\cite{2005ApJ...618..195G} we typically reach more than twice as large
cluster-centric distances, while we go a couple of magnitude deeper
than \cite{2007AJ....134..466K}.

Using $g$, $r$ and $i$ photometry we were also able to measure the
colors of the ICL and found that they are completely consistent with
those derived for the total light of the cluster, thus implying that
the ICL is emitted by stellar populations that do not differ
significantly from those present in galaxies.

\subsection{Alignment of the ICL with the BCG}
An important test to understand the origin of the ICL is to check its
alignment with other cluster components and with the BCG in
particular. The stacking of randomly oriented images naturally
produces a radially symmetric SB distribution. However, if one selects
only clusters with significantly elongated BCG and aligns the images
along its major axis, the symmetry is broken and it is possible to
study the mutual alignment and elongation of the different components.
We find that the stacked ICL is more elongated than the BCG, while the
distribution of galaxies is less elongated. This implies that the ICL
is well aligned with the BCG and is also more elongated on average.
The distribution of galaxies, as opposed, is either less elongated or
worse aligned with the BCG, or both.

\subsection{The ICL integral flux}
Measuring the total flux of the ICL requires to subtract the
contribution of the BCG from the diffuse component. Unfortunately
there is no physically motivated method that can be adopted for this
task and, as I will discuss below, different authors adopt different
strategies. We decided to measure the flux contributed by the BCG as
the integral of the DeVaucouleurs inner component (as shown in Fig.
\ref{fig:r_profile}). The second problem that arises in this measure is
the aperture (if any) within which the fluxes are measured. Given the
large uncertainties related to any extrapolation, we prefer to be
conservative and limit our integration to within 500 kpc, where we
have reliable data and uncertainties in the background subtraction are
negligible. As it will be discussed below, a fixed metric aperture
might not be the best choice when clusters of different masses are
compared. However, given the large uncertainties of $R_{200}$ for our
clusters, a fixed metric aperture is the safest solution.

We find that $(10.9\pm5)\%$ of the optical luminosity of a
cluster within 500 kpc has to be attributed to the ICL, another
$(21.9\pm3)\%$ is contributed by the inner DeVaucouleurs component of the
BCG. These fractions appear to have very little dependence on cluster
richness or BCG luminosity.

\subsection{Dependence on cluster properties}
\begin{figure}
\centerline{\includegraphics[width=0.75\textwidth]{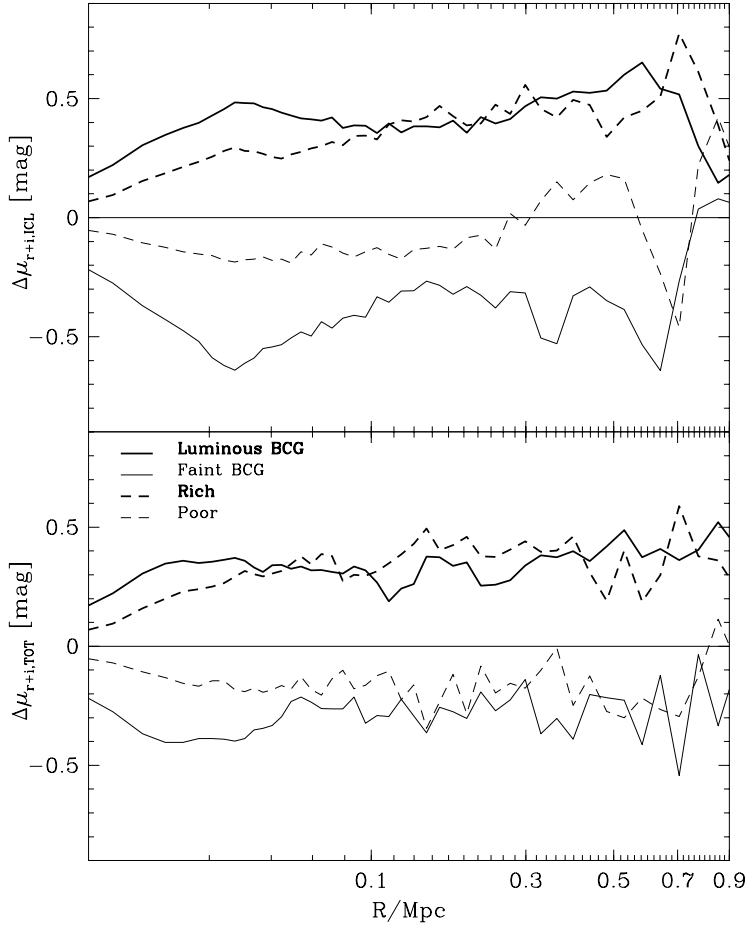}}
\caption{Comparisons between different subsamples: SB difference
  between the subsamples and the main sample. Thick solid lines are
  used for clusters hosting a luminous BCG, thin solid for faint BCG,
  dashed thick for rich clusters, and dashed thin for poor clusters.
  The bottom panel displays the SB differences as a function of the
  radius for the {\it total} light, the top panel those for the
  corrected {\it ICL+BCG} component alone.}\label{fig:trends}
\end{figure}

The broad range in richness and BCG luminosity enables us to study the
dependence of the ICL brightness and flux upon those cluster
properties. To this purpose we consider the clusters in the bottom and
top tertiles of the richness distribution and similarly for the
distribution in BCG luminosity. We end up with subsamples of ``rich''
clusters (estimated mass $M\gtrsim5~10^{13}\mathrm{M_\odot}$, as in
Sec. \ref{sec:sample_photo}), ``poor'' clusters
($M\lesssim3.5~10^{13}\mathrm{M_\odot}$), clusters with a ``luminous''
BCG ($M_{r,0}<-23.25$~mag) and clusters with a ``faint'' BCG
($M_{r,0}>-22.85$~mag). We then compare the SB profiles of the total
and diffuse light for these four subsamples versus the average SB
profiles of the complete sample, as shown in Fig. \ref{fig:trends}.
The total light follows expected trends, with ``rich'' and
``bright-BCG'' clusters being brighter than the other two classes. On
the other hand, the diffuse light is particularly suppressed in
``faint-BCG'' clusters only, while ``poor'' clusters do not show such
a strong effect. This suggests that the presence of a bright central
galaxy is a crucial ingredient for the formation of the ICL.

\section{Interpretation}\label{sec:interpretation}
The results presented in the previous section lead to conclude that,
although the ICL cannot be considered just as an extension of the BCG,
there are strong links between these two components. The ICL tends to
concentrate around the BCG; also, there must be some link between the
growth of these two components since we found that the lack of a
bright BCG implies a strong suppression in the ICL.  It appears that
most of the action in creating the ICL occurs close to the bottom of
the cluster potential well, where galaxies experience very strong
tidal fields and may eventually merge into the BCG. During these
events a large number of stars becomes unbound as shown by numerical
simulations \citep[see for
instance][]{2007MNRAS.377....2M}. Interestingly, the high degree of
alignment and elongation of the ICL with respect to elongated BCGs
suggests that the accretion of galaxies along radial orbits
(filaments) is particularly efficient in generating intracluster
stars.

\section{Open issues}\label{sec:open}
The amount of ICL relative to the total cluster luminosity and the
luminosity of the BCG is the quantity that is most often used to test
against model predictions. Yet there are still many statistical
uncertainties and bias in this measurement (and in the corresponding
model predictions) that severely hamper a clean comparison. In
particular, I discuss two very relevant aspects of this issue here
below.
\subsection{ICL vs BCG vs Halos}
As already mentioned in the Introduction, pure photometric
measurements of the ICL have to adopt operative definitions. In
practice SB thresholds are introduced, below which the light is
considered as ICL. However galaxies do not have sharp boundaries, but
rather have smoothly declining SB profiles and are possibly surrounded
by low-SB halos. The diffuse light which is measured in photometric
studies is contaminated by these low-SB structures. Different choices
of SB thresholds should be taken into account when comparing different
studies, although the exact value of these thresholds do not appear to
be critical \citep[see][their Sec. 5.4]{SZ+05}. On the other hand,
the operational definition of ICL in terms of SB might lead to
substantially biased results when comparing with simulations that
define the ICL according to dynamical criteria.  Recent simulations
\citep[e.g.][]{2006ApJ...648..936R,2007astro.ph..3374C} mimic the
definition of ICL in terms of SB and ease the comparison with the
observations. It is unclear though how the dynamical and photometric
ICL compare to each other and how much of the discrepancy between the
predictions of different simulations arises from the very definition
of ICL.

Separating the ICL from the BCG and its halo is another very
controversial point, both in observations and simulation. The point is
well illustrated by a comparison between \cite{2005ApJ...618..195G}
and \cite{SZ+05}, who have the largest samples in literature and
explicitly analyze the relationship between ICL and BCG.
\cite{2005ApJ...618..195G} decompose the SB distribution of BCG+ICL in
two DeVaucouleurs components, of which the outer one is consider as
genuine ICL and the inner one as proper BCG. \cite{SZ+05} instead fit
only the inner DeVaucouleurs profile and consider it as BCG, while all
the residuals are considered as ICL. \cite{2005ApJ...618..195G} give
$ICL/BCG>5$ while \cite{SZ+05} find $ICL/BCG<0.5$. Many factors may
play a role in creating this discrepancy, for instance the different
extent of the photometry and a different sample selection (the
Gonzalez et al. sample, in fact, is biased toward clusters with a
luminous dominant BCG). However, we calculate that by applying a
2-DeVaucouleurs decomposition to the \cite{SZ+05} data one would
obtain $ICL/BCG\sim2$, hence much closer to the
\cite{2005ApJ...618..195G} result. We conclude that the ratio of
ICL/BCG is very poorly constrained by present observations and it
would better not be used for comparison with models. As opposed, the
ratio (ICL+BCG)/total is much more robust and there is apparently
general consensus on a value around 30\%.

\subsection{Trends with cluster mass and richness}
The literature about ICL of recent years presents a broad variety of
claims about the dependence of the ICL fraction upon cluster mass or
richness, including some contrasting claims of trends. As an
illustration of the current state of the observations I report the
trends of the ICL fraction in different works that analyze a broad
range of cluster properties:
\begin{itemize}
\item{\cite{SZ+05}: no trend with richness nor with BCG luminosity
    (but brightness correlates with BCG luminosity).}
\item{\cite{2007AJ....134..466K}: no trend with mass, anti-correlation
    with presence of a cD.}
\item{\cite{2007arXiv0705.1726G}: negative trend with mass.}
\end{itemize}

It is worth noting that the trends (or lack of trends) reported by
\cite{SZ+05} could be intrinsically biased by the adoption of a fixed
metric aperture of 500 kpc, which correspond to smaller fraction of
$R_{200}$ for more massive clusters. Given the steeper profile of the
ICL with respect to galaxies, the ICL fraction of more massive
clusters could be overestimated and a correction for this effect could
reconcile these results with the negative trend found by
\cite{2007arXiv0705.1726G}.

From the short and incomplete compilation reported above it is evident
that better determinations of the trends of the ICL with cluster mass
and richness and with the luminosity of the BCG and the presence of a
cD in the cluster are needed. These trends appear to be distinctive
and crucial for theoretical models too. A better observational
determination will certainly gain better insights in the dynamics of
cluster formation.

\section{Conclusions}\label{sec:conclusions}
I have reviewed the main results on the ICL from the stacking analysis
performed by \cite{SZ+05}. Thanks to the high quality of the SDSS
drift-scan data, the large sample (683 clusters) and the effectiveness
of the stacking technique against systematics, these results represent
the most robust \emph{statistical} assessment of the properties of the
ICL so far and are an essential complement to detailed photometric and
kinematic studies of the ICL in individual clusters.

I presented a tentative physical interpretation which is consistent
with models where most of the ICL is created during violent tidal
interactions in the deepest regions of a cluster's potential; these
processes appear to be linked to the growth of the BCG.

We are witnessing a general convergence, within factor 2 or better for
the basic observables, between different observations and models of
the ICL. This is somehow astonishing, given the extreme difficulties
of the measurements. However many open and controversial issues are
still present. I indicate in particular the problems related to the
operational definition of ICL luminosity and fractions, and to the
assessment of trends with cluster global properties as priorities for
the progress in this field.

%\begin{acknowledgments}
%We would like to acknowledge the useful comments of a referee concerning
%the solution procedure used in \S\,\ref{sec:concl}. A.\,N.\,O. is supported
%by SERC under grant number GR/F/12345.
%\end{acknowledgments}

\end{document}